# Gamma-ray Strength Functions in Thermally Excited Rotating Nuclei*

T. Døssing, B. Herskind

The Niels Bohr Institute, Blegdamsvej 17, DK2100 Copenhagen Ø, Denmark

A. Maj

Niewodniczanski Institute of Nuclear Physics, Krakow, Poland

M. Matsuo

Graduate School of Science and Technology, Niigata University, Niigata 950-2181, Japan

E. Vigezzi, A. Bracco, S. Leoni, R.A. Broglia

INFN sez Milano, and Department of Physics, University of Milano, Milano 20133, Italy

A general discussion and illustration is given of strength functions for rotational transitions in two-dimensional $E_{\gamma 1} \times E_{\gamma 2}$ spectra. Especially, a narrow component should be proportional to the compound damping width, related to the mixing of basis rotational bands into compound bands with fragmented transition strength. Three $E_{\gamma 1} \times E_{\gamma 2}$ spectra are made by setting gates on triple coincidences, selecting cascades which feed into specific low-lying bands in the nucleus $^{168}$Hf. In each of the gated spectra, we find a ridge, carrying about 100 decay paths. This ridge is ascribed to rotational transitions in the excitation energy range of 1.0 to 1.5 MeV above the yrast line. The FWHM of the ridges are around 40 keV, about a factor of two smaller than calculated on the basis of mixed cranked mean field bands.

---

* Presented at the High Spin Physics 2001 Conference, Warsaw, Poland, February 6–10, 2001.

### 1. Introduction

(HI,xn) reactions produce compound nuclei within an "entry region" of energy and angular momentum, with a typical energy above yrast of half the neutron separation energy. ¿From the entry region reached by the last particle evaporation, $\gamma$-cascades pass through thermally excited states on their way towards the cold yrast line and eventually the ground state at temperature $T = 0$. To learn about the rotation of thermally excited states, one applies a combination of discrete spectroscopy and spectroscopy of unresolved $\gamma$-energy transitions.

Figure 1 illustrates the $\gamma$-rays emitted in three steps of a cascade starting either from cold states at temperature $T \sim 0$, or from a warm state at temperature $T > 0$.

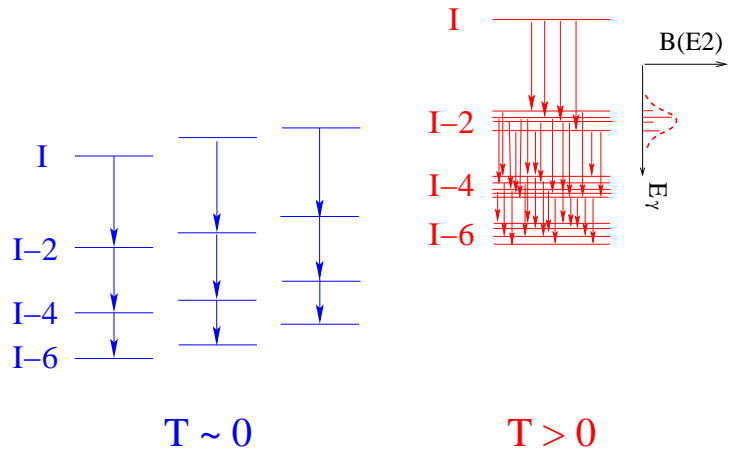

Fig. 1. Illustration of three steps in a rotational cascade, for cold bands (left hand side) or starting from a mixed band state (right hand side). To the far right is illustrated transition strengths for the actual state (bars), and after averaging over many initial states, smooth dashed curve.

This conference has provided many examples of the rewards of resolving discrete lines for cold states, such as band terminations related to specific nucleonic orbitals, E1 and M1 bands related to the symmetries and internal structure, magnetic rotation and wobbling motion, to mention a few.

Concerning the warm rotation, $T > 0$, typically 100 initial states at each angular momentum are populated by the cascades. Considering the branching on several states by the cascades in each step, it would be a formidable task to piece together a complete level scheme. With so many states and transitions, probably less significance may be attached to the in-dividual states, but rather to their average properties. In the present paper, we shall especially consider strength functions for the rotational quadrupole decay, both distributions averaged over many states, as well as fluctuations of strengths. The mixing of rotational bands and fragmentation of rotational strength is caused by the interactions among rotational bands. These interactions, and their implications for the rotational decay, shed light on properties of the rotational bands, which cannot be studied for cold bands. Especially interesting is the transition from regular rotational bands to chaotic rotation with fragmented rotational transition strengths with increasing temperature.

The fragmentation of rotational strength was first discussed by G. Leander [1], motivated by the rather weak rotational correlations observed in two-dimensional spectra, (sometimes also called energy-energy correlation spectra [2, 3]). A schematic model of mixed rotational bands and the associated fragmentation of the rotational strength was proposed by B. Lauritzen et al. [4]. Here, the basis bands are taken as cranked mean field bands, interacting via a two-body force. Also, emphasis was put on the generic character of the type of rotational motion carried by the mixed bands of various frequencies, independent on the details of the nature of the bands and their interaction. To stress the analogy to other nuclear and condensed matter phenomena, this motion was named "damping of rotational motion" [5]. More realistic calculations, emphasizing general order to chaos properties of states and strengths with increasing temperature, were carried out by S. Åberg [6, 7]. The most consistent and specific calculations of mixed rotational bands have sofar been carried out by M. Matsuo et. al. [8, 9].

These calculations are based on the cranked mean field, and the diagonalization of the residual two-body force is performed within a rather restricted basis. The basis gives a properly exponentially increasing average level density, including many-particle-many-hole states. However, the excitation in the basis of the individual particle away from the Fermi surface is effectively limited to the maximum temperature in the basis of about 0.6 MeV, corresponding to the lowest about 3000 states of each parity and signature at each angular momentum. The use of the cranked mean field implies that the rotational E2 matrix elements are assigned to the basis bands as an ansatz. The restriction to modest nucleon excitations away from the Fermi surface implies that phase correlations of matrix elements of the residual interaction will not generate coherent states, that is in the present case coherent combinations of bands. This is especially a problem for the paring part of the residual interaction, but also surface vibrations are to a high degree neglected by this procedure.

In the present paper, we will apply a combination of investigating the shape [10, 11] and the fluctuations [12, 13] of two-dimensional spectra. This investigation is concerned with the distribution of $\gamma$-ray energies in two consecutive steps of a cascade, that is the two-step strength function as it will appear in a two dimensional $E_{\gamma 1} \times E_{\gamma 2}$ plane.

## 2. $\Delta I = 0, -2, -4$ strength functions

The strength functions underlying damped rotational motion are illustrated in Figure 2.

### 2.1. $\Delta I = 0$; Compound damping width

Locally, at each angular momentum, the basis bands, which carry the rotational strength, are mixed by an interaction. In the model by Matsuo et. al. [8], the bands are taken as mean-field states of a rotating potential, that is they are n-particle-n-hole states relative the the yrast state, $n = 0, 1, 2, 3, \cdots$. Each mean field state $|\mu\rangle$ is spread over an energy interval $\Gamma_\mu$ by the residual interaction.

The quantity $\Gamma_\mu$ carries important information on the nature of the basis bands, and their interaction with surrounding states. This interaction leads to the formation of compound states as superpositions of basis states, and a proper name for $\Gamma_\mu$ is *the compound damping width*.

### 2.2. $\Delta I = -2$; Rotational damping width

The mixed band states carry rotational strength, relating back to their content of basis band states. Rotational transitions compare the mixing at two neighboring angular momenta, $I$ and $I - 2$, and the B(E2) connecting such two states is given by

$$B(E2)(\alpha(I) \to \alpha'(I-2)) = B(E2)_{b.b.} \left| \sum_\mu \langle \alpha(I) | \mu(I) \rangle \langle \mu(I-2) | \alpha'(I-2) \rangle \right|^2$$

where $B(E2)_{b.b.}$ denotes the rotational matrix element of basis bands. Averaging over an energy interval of compound $|\alpha\rangle$ states, and summing over final states $|\alpha'\rangle$, one finds a rotational strength function, centered around some average energy, and dispersed over an energy range of FWHM defined as *the rotational damping width*. For not too high excitation energy above yrast, the rotational damping width will be twice the width of the distribution of rotational frequencies of the basis bands mixed into the compound states.

For the individual $|\alpha\rangle$-state, specific strong $|\mu\rangle$ components will give additional structure to the strength function. One strong $|\mu\rangle$ component will place some intense transitions within an interval of width $\Gamma_\mu$, centered around transition energy $2\omega_\mu$.

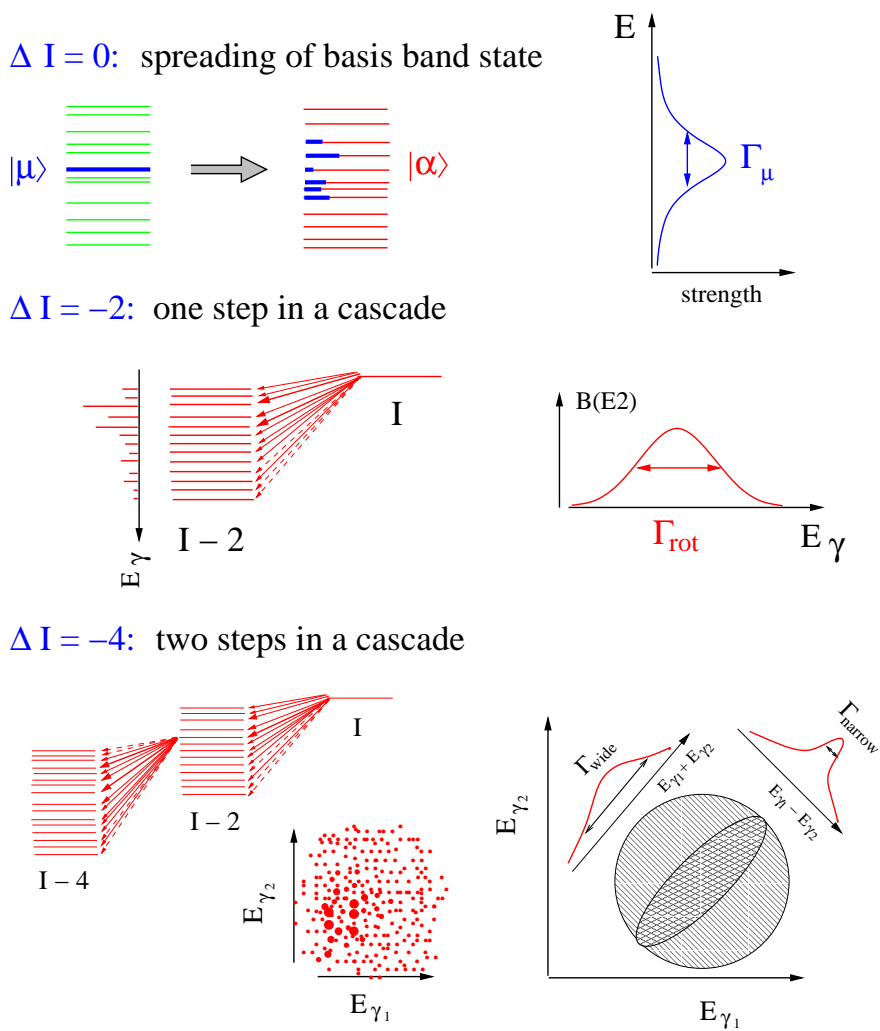

Fig. 2. Illustration of the relevant strength functions which underly the wide and narrow components of two consecutive transitions of a cascade. On the left hand part of each panel is shown individual states mixing into compound states (for $\Delta I = 0$), or transitions from individual states (for $\Delta I = -2$ and $\Delta I = -4$). On the right hand part of each panel is shown the resulting strength function averaged over many states.

### 2.3. $\Delta I = -4$; Wide and narrow component

Sofar, the strength function for $\Delta I = -2$ for individual mixed states cannot be isolated and observed. But the point is now that the next step in the decay from the states $|\alpha'(I-2)\rangle$ to $|\alpha''(I-4)\rangle$ also picks out the same strong $|\mu\rangle$ components. For two steps in the decay out from an individual $|\alpha(I)\rangle$, (illustrated by the dots on the middle lower panel of the figure), this will result in a two-component strength function of the two $\gamma$-ray energies. Intense transitions are concentrated around the transitions energies that would be along the $|\mu\rangle$ basis band, smeared out by the compound damping width $\Gamma_\mu$ in both directions. For the different $|\alpha\rangle$ states, the $|\mu\rangle$ states strongly present will sometimes have smaller rotational frequency than the average, (the example picked for the figure, and displayed in the middle of the lower panel), sometimes larger. Averaging over an interval in energy of mixed states, a two-component strength function evolves which has a wide approximately circular part, and a part which is narrow in the $E_{\gamma 1} - E_{\gamma 2}$ direction, as illustrated on the right lower corner of figure 3. ¿From this discussion, the widths of the wide and narrow components are expected to be of the order of:

$$\Gamma_{wide} \approx 2\Gamma_{rot} , \qquad \Gamma_{narrow} \approx 2\Gamma_\mu$$

Going up in temperature, the narrow component will be carried by the Porter-Thomas strength fluctuations of the mixing matrix elements $\langle \mu | \alpha \rangle$, and on this basis, the narrow component can be estimated to have the intensity

$$(Intens.)_{narrow} \sim \frac{1}{\rho \Gamma_\mu}$$

where $\rho$ is the level density. Figures 1 and 4 of ref. [14] display specific mixed band calculations of these quantities, in qualitative accordance with the present illustrations.

Going still further up in temperature, the compound damping width $\Gamma_\mu$ is expected to become larger than the rotational damping width, and the narrow component will dissolve into the wide. The temperature for this dissolving is also the temperature for the onset of motional narrowing of the rotational damping [4]. However, with the estimates based on the cranking model, this will only be at the high-energy edge of the states populated by the gamma-cascades for rare earth nuclei via fusion reactions. For superdeformed nuclei, motional narrowing of the rotational damping may be more easily accessible [9, 15].

## 3. Experimental search for narrow component

### 3.1. Number of bands as function of irregularity parameter $\delta$

A search for the narrow component is reported in reference [16], applying three-dimensional spectra, $E_{\gamma 1} \times E_{\gamma 2} \times E_{\gamma 3}$. For three consecutive $\gamma$-rays,

$E_{\gamma 1}$, $E_{\gamma 3}$, $E_{\gamma 2}$, the quantity

$$\delta = E_{\gamma 1} - 2E_{\gamma 3} + E_{\gamma 2}$$

will be zero for a perfectly regular rotational band, whose energy as function of angular momentum contains terms up to second order when approximated around an angular momentum $I_0$, appropriate for the transition energy interval under investigation. Deviations from $\delta = 0$ may be due to gradual alignment of orbitals, gradual changes in the pairing properties, or to shape changes along the band. More important and dramatic are the irregularities induced by band crossings, which produce coincidences with specific values of $\delta$, depending on the relative slopes of the bands, and the magnitude of the residual interaction at the band crossing.

In ref. [16], the ridge in $E_{\gamma 1} - E_{\gamma 2}$ for different gates on irregularity $\delta$ was investigated. It was found that with increasing $|\delta|$, the number of paths on the ridge first increases, then decreases, with the average intensity of each transition on the ridge decreasing. Thus, the bands most intensely populated by the cascades close to yrast are the most regular in energy. With increasing excitation energy up from yrast, the bands become more irregular. The width of the ridge found by this technique is of the order of 30 keV, and we loosely estimate that an excitation energy region of about 0.6 to 1.0 MeV is probed. Extrapolating this behavior in a somewhat speculative way, we think that this irregularity is a precursor to the narrow component, which should develop with still increasing temperature and beginning fragmentation of the strength.

### 3.2. Two-dimensional spectra gated by specific bands

At present, we apply another technique, namely to investigate the shape and fluctuations of the first ridge of spectra gated by specific low-lying rotational bands. The idea behind this technique is illustrated on figure 3. If the E1 transitions feeding into the band mainly come from the region of the flow-line, the two last rotational transitions preceding the gating transition, should display a strength function as the one illustrated on figure 2, lower right corner.

The idea is tested on a high-statistics experiment performed with gammasphere at ANL using a $^{76}$Ge + $^{96}$Zr fusion reaction to populate the highest spin region in $^{168}$Hf nuclei. (see ref's [17] and [18] for experimental details). The three most intense rotational bands in $^{168}$Hf are selected by the gates given in table 1.

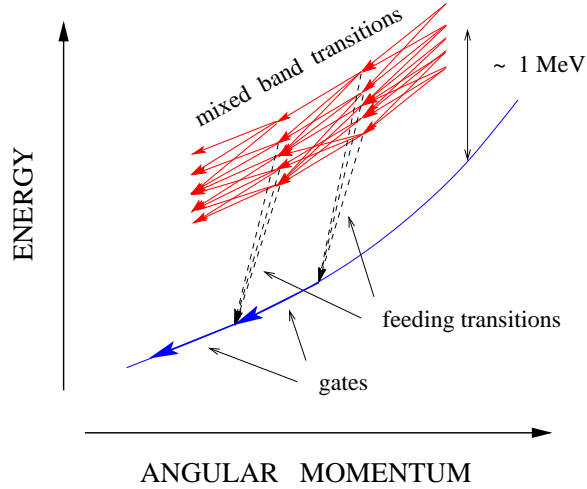

Fig. 3. Illustration of feeding from the flow line down into transitions on a gating band. The last two rotational steps preceding the feeding down will display a strength function as illustrated on figure 2, lower right corner.

| $^{168}$Hf bands | AB | AE | AF |
|---|---|---|---|
| $E_\gamma$ | 607 keV | 588 keV | 559 keV |
| $I^\pi \to I - 2^\pi$ | $20^+ \to 18^+$ | $19^- \to 17^-$ | $18^- \to 16^-$ |
| | 684 keV | 619 keV | 598 keV |
| | $22^+ \to 20^+$ | $21^- \to 19^-$ | $20^- \to 18^-$ |
| | 751 keV | 781 keV | 694 keV |
| | $24^+ \to 22^+$ | $27^- \to 25^-$ | $24^- \to 22^-$ |
| | 812 keV | 851 keV | 831 keV |
| | $26^+ \to 24^+$ | $29^- \to 27^-$ | $28^- \to 26^-$ |

Table 1. The energy in keV and the angular momenta of the gating transitions for the three intense bands in $^{168}$Hf. The standard labels are applied for neutron quasiparticles in deformed rotating potentials with pairing.

The first and second moments of counts $\mu_1, \mu_2$ [12] are evaluated for all the gated $E_1 \times E_2$ spectra.

On the planes, clear ridges are observed. So-called "perpendicular cuts" covering intervals in $(E_{\gamma 1} + E_{\gamma 2})/2$, as function of $E_{\gamma 1} - E_{\gamma 2}$ are made, and displayed in figure 4 for the AE gate. Alternating, the cuts contain (for example for 880 keV, the second curve from the bottom on both panels), or exclude (for example for 912 keV, the third curve) coincidences on the gating band on the first ridge. The first ridge is situated around $E_{\gamma 1} - E_{\gamma 2} \approx \frac{4}{\mathcal{I}} \sim 60$ keV, containing consecutive transitions along bands or belonging to the narrow component ($\mathcal{I}$ is a typical moment of inertia). One sees that the coincidences on the gating band above the gates makes the ridge very sharp and strongly fluctuating. This is due to the one sharp spike in the spectrum.

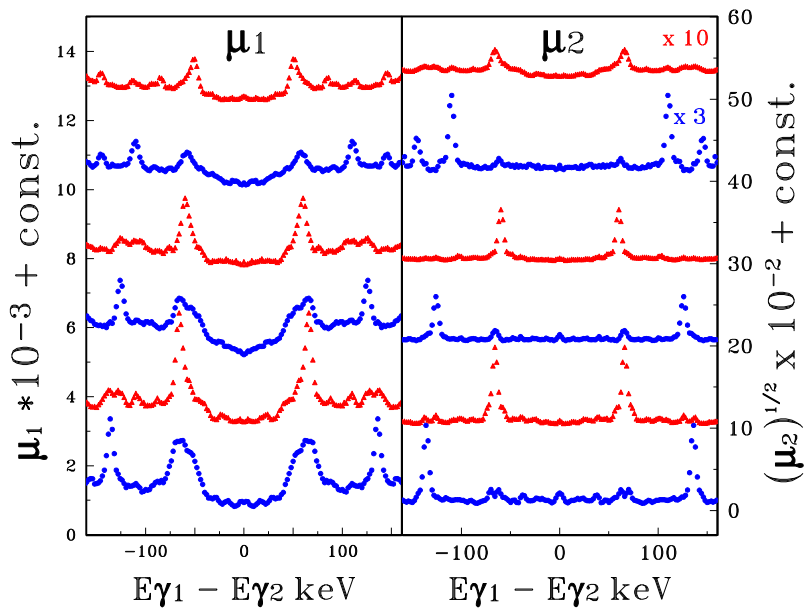

Fig. 4. Perpendicular cuts of the first moment of two-dimensional spectra, gated by the band AE of $^{168}$Hf, left hand side, and the second moment, measuring fluctuations of counts, right hand side. From the raw gated spectra, an "UNCOR" product spectrum has been subtracted, and the first moment implies a smoothing of the resulting "COR" spectrum, with a narrow Gaussian smoothing function (of width equal to one channel). The curves are displayed with a shift on each curve, and they contain projections on the $E_{\gamma 1} - E_{\gamma 2}$ axis covering intervals in the coordinate $(E_{\gamma 1} + E_{\gamma 2})/2$, which are 21 channels (42 keV) wide and centered around $(E_{\gamma 1} + E_{\gamma 2})/2 = 850, 882, 916, 944, 976$ and 1000 keV.

A clear systematics of alternating results are found on figure 4, and one can go on to evaluate the number of paths, and the width of the ridge, shown on figure 5. With the band included, the number of paths becomes about 4, and without the band, the number of paths is typically 100. This last number is substantially larger than any previous ridge number of paths observed [12, 19, 13], and give a qualitative confirmation of the picture of feeding illustrated on figure 3: The side-feeding into the band occurs mostly as direct transitions down from the main flow of cascades through excited states above yrast, and not so much through cross-over transitions from other cold bands.

More quantitatively, the lowest 100 states from yrast and up of all four parity-signature combinations cover approximately the lowest 1.1 MeV of energy, applying standard level density estimates [20]. Taking into account selection rules in the feeding transitions, one would need to go to even higher energies above yrast, but this is counteracted, when one also considers the fragmentation of the rotational strength in each step. Altogether, we estimate that the large number of paths shows that the average energy of states probed by the wide ridges cannot be smaller than 1 MeV, and not larger than 1.5 MeV. The width of the wide ridges, $\Gamma \approx 30-40$ keV, is about a factor of two smaller than estimates of the narrow component based on cranked mixed bands [8, 14].

Going up in transition energy, the increasing energy of the states at the flow line should lead to an increase in both the number of paths and the width of the wide ridges. However, the opposite is observed. At transition energy around 1000 keV, corresponding to angular momentum around 30-34$\hbar$, the ridge is only 20 keV wide. Transitions with $E_\gamma \sim 1000$ keV precede the gating transitions by about 5-6 steps of cascades. Such cascades could pass through bands with high alignments, which are closer to yrast around 30-34 $\hbar$, but which go up from yrast 5 steps lower down in angular momentum. They then eventually become mixed, and can feed down into the gating bands.

## 4. Concluding remarks

The improvements in counting statistics achieved with the newest detection arrays allows one to probe excited rotational bands in more detail. For quite a number of nuclei, specifically $^{163}$Lu, $^{163}$Er and $^{168}$Hf, typically $10^6$ consecutive three-step transitions of $\gamma$-cascades are recorded, for example going through the angular momenta $34 \to 32 \to 30 \to 28$. So many counts are needed for the fluctuations of the spectra to rise above those trivially given by the counting statistics.

Gating on specific rotational bands in $^{168}$Hf, one sees a ridge containing about 100 paths, corresponding to transitions from an energy interval above yrast of somewhere between 1.0 and 1.5 MeV. This ridge is interpreted as the narrow component, relating back the compound damping width of basis bands. The width of the component, about 35 keV, is smaller than mixed cranking model estimates by about a factor of two.

In the near future, the gating conditions will be simulated in cascade simulations, to make a more detailed comparison with the mixed band calculations.

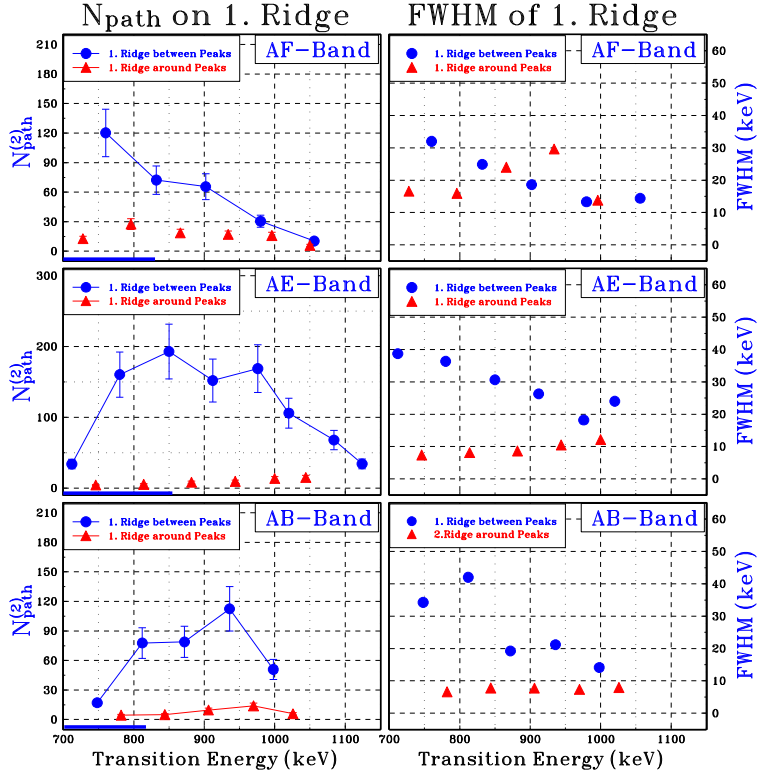

Fig. 5. Number of paths and the FWHM of the ridges displayed on figure 4 are shown as function of $\gamma$-ray energy, for the three different bands, with gates summed over the transitions given in table 1.

By evaluating covariances [13] between spectra set by different gates, one can investigate selection rules for the K-quantum number [21], or the shape, how such selection rules are obeyed by the cascades flowing through different regions in temperature and angular frequency. A promising data set for $^{163}$Lu displays two distinct first ridges, characterized by two different moments of inertia, corresponding to normally deformed and triaxial superdeformed shapes [22]. Fluctuations and covariances on these ridges will tell about the narrow components, and also the amount of interaction at band crossings between the excited bands of the two shapes.

**Acknowledgement:** A.M. acknowledges a partial support from Polish State Comittee for Scientific Research (KBN) under Grant No. 2 P03B 001 16.